\documentclass{article}
\usepackage{PRIMEarxiv}

\usepackage[utf8]{inputenc} 
\usepackage[T1]{fontenc}    
\usepackage{hyperref}       
\usepackage{url}            
\usepackage{booktabs}       
\usepackage{amsfonts}       
\usepackage{nicefrac}       
\usepackage{microtype}      
\usepackage{lipsum}
\usepackage{fancyhdr}       
\usepackage{graphicx}       
\graphicspath{{media/}}     
\usepackage{xcolor}
\usepackage{amssymb}
\usepackage{amsmath}
\usepackage{array}
\usepackage[figuresright]{rotating}
\usepackage{enumitem}
\usepackage{cleveref}
\usepackage{diagbox}
\usepackage{float}

\title{FinArena: A Human-Agent Collaboration Framework for Financial Market Analysis and Forecasting}

\author{
  Congluo Xu \\
  Business School\\
  Sichuan University\\
  Sichuan, China \\
  \texttt{xucongluo@stu.scu.edu.cn} \\
   \And
  Zhaobin Liu \\
  Department of Information Systems\\
  City University of Hong Kong\\
  Hong Kong, China \\
  \texttt{zhaobiliu2-c@my.cityu.edu.hk} \\
   \And
  Ziyang Li\thanks{Corresponding Author} \\
  Business School\\
  Sichuan University\\
  Sichuan, China \\
  \texttt{lzy\_feng@scu.edu.cn}
}

\begin{document}
\maketitle

\begin{abstract}
To improve stock trend predictions and support personalized investment decisions, this paper proposes FinArena, a novel Human-Agent collaboration framework. Inspired by the mixture of experts (MoE) approach, FinArena combines multimodal financial data analysis with user interaction. The human module features an interactive interface that captures individual risk preferences, allowing personalized investment strategies. The machine module utilizes a Large Language Model-based (LLM-based) multi-agent system to integrate diverse data sources, such as stock prices, news articles, and financial statements. To address hallucinations in LLMs, FinArena employs the adaptive Retrieval-Augmented Generative (RAG) method for processing unstructured news data. Finally, a universal expert agent makes investment decisions based on the features extracted from multimodal data and investors' individual risk preferences. Extensive experiments show that FinArena surpasses both traditional and state-of-the-art benchmarks in stock trend prediction and yields promising results in trading simulations across various risk profiles. These findings highlight FinArena’s potential to enhance investment outcomes by aligning strategic insights with personalized risk considerations.
\end{abstract}

\keywords{Human-Agent Collaboration \and Large Language Models \and Stock Forecasting \and Adaptive Retrieval-Augmented Generation (RAG) }

\section{Introduction}
\label{Section 1}

Financial markets have long been considered as inherently complex and dynamic systems \cite{MOUCK1998189, Lux1999ScalingAC}, characterized by intricate inter-dependencies among economic indicators, geopolitical events, and investor behaviors. These systems experience rapid fluctuations influenced by a myriad of factors, making accurate analysis and forecasting increasingly challenging. Traditional financial analysis methods often rely on linear models and one-dimensional data, while many researches have discovered long-term and short-term nonlinearity in financial markets \cite{wang2020nonlinear,sadique2001long}. Those statistical approaches such as CAPM \cite{PORTFOLIO}, Multi-factor Asset Pricing models \cite{fama2015five},  Autoregressive Integrated Moving Average model (ARIMA) \cite{Box1978TimeSA}, struggle to capture the nonlinear characteristics. Moreover, unstructured data such as news articles, global economic reports and announcement statements have a wealth of available information that can be used to aid forecasting and decision-making \cite{KHADJEHNASSIRTOUSSI20147653, 10.1145/3533018},  but traditional methods are unable to consider and leverage it as well.

The emergence of Large Language Models (LLMs) has been transforming AI with their exceptional performance and generalization across tasks, while also showing significant potential for applications in financial analysis. Depending on advanced natural language processing capabilities, LLMs such as OpenAI's GPT series \cite{gpt1,gpt2,DBLP:journals/corr/abs-2005-14165,openai2024gpt4technicalreport} can interpret and generate human-like text, enabling quantitative analysis of financial news and the extraction of sentiment from diverse sources. They help human analysts by providing insights derived from vast datasets, identifying emerging trends, and aiding in the formulation of investment strategies. Some research efforts have tailored LLMs with domain-specific knowledge in finance, exemplified by models such as FinBERT \cite{FINBERT19,FINBERT20,ijcai2020p622} and BloombergGPT \cite{bloomberggpt}, showing good performance on financial sentiment analysis, entity recognition tasks, and other complex finance tasks.

Initial applications of LLMs in finance have revealed their ability to efficiently process unstructured data, assisting financial decision-making and supporting more informed market predictions.

An overall performance test for LLMs on multiple financial tasks \cite{Zhao2024RevolutionizingFW} reveals that LLMs can effectively follow instructions to complete tasks on these various tasks. Another research achieves the integration of transaction data, technical analysis and sentiment scores for trend prediction of price \cite{zhou2023adaptive}, and the results in \cite{Li2023LargeLM}  are evaluated in terms of language tasks, sentiment analysis, financial reasoning, financial Q \& A, and the behavior simulation for participants in the financial market. These investigations demonstrate the potential of LLM in the financial field.

However, the implementation of LLMs for financial analysis faces three major challenges.
\textbf{First}, LLMs often struggle with uncertainty when confronting relevant information that was not present in their pre-training data \cite{xiong2024llmsexpressuncertaintyempirical}. This limitation can lead to hallucinations, where the model generates inaccurate or misleading information. In the fast-paced financial domain, new events and data emerge continuously, and relying on outdated training data can result in significant analytical errors and unreliable predictions. The inability of LLMs to incorporate the latest information undermines their effectiveness in providing accurate assessments and forecasts.
\textbf{Second}, financial market analysis requires the integration of diverse multimodal data sources, including time-series data, tabular information from financial statements, and unstructured data such as news sentiment. Traditional time-series models typically focus on a single modality, which constrains their ability to fully utilize the breadth of available information. LLMs, on the other hand, primarily model language by predicting subsequent tokens in a sequence. Some recent studies have sought to apply LLMs to time-series data by framing the problem as a token sequence modeling task, thereby employing LLMs for pattern recognition and reasoning. However, the effectiveness of this approach remains a subject of debate, given the distinct characteristics of time-series data compared to natural language \cite{tan2024are}. More recently, inspired by advanced large models such as GPT-4, researchers have developed large transformer models trained independently on extensive time-series datasets to minimize prediction errors \cite{jin2024timellm, Cao2023TEMPOPG}. Despite the progress made with different foundation models that excel in specific modalities, the challenge of effectively integrating these modalities and utilizing LLMs for comprehensive financial analysis remains an area in need of further in-depth investigation.
\textbf{Third}, most of the existing researches focus on the theme of ``Human-Machine confrontation''\cite{CAO2024103910}, that is, proposing various machine learning methods, multi-LLM frameworks or fine-tuning technologies to ``defeat'' top human experts, but few researches focus on how to deploy Human-Agent collaboration framework to improve investment profitability for ordinary investors, which should be the original intention of human beings to design these analysis tools.

To address these challenges, this study proposes a Human-Agent collaboration framework for handling multimodal financial data analysis and investor’s risk preference input, called FinArena. Specifically, the proposed framework is inspired by a mixture of experts (MoE) approach \cite{Jacobs_Jordan_Nowlan_Hinton_1991}.
In human module, FinArena incorporates an interactive interface that captures individual risk preferences, enabling personalized investment strategies. In machine module, FinArena integrates multiple LLMs, specializing in different types of financial data, such as historical time series of stock price, corporate news, and financial statement report information. Each expert group is equipped with one or more LLMs, conducting independent data analysis and processing for their specific type of information. Finally, a universal expert model synthesizes these analyzes, incorporating the individual risk preferences of investors during its analysis, thereby mimicking the division of labor in a real-world investment team. Additionally, for news data, we not only employ a self-reflection process during pre-processing to actively filter out irrelevant information, but also introduce an adaptive Retrieval-Augmented Generation (RAG) method \cite{10.5555/3495724.3496517} during formal processing. This allows the model to access additional information through Internet searches if needed, significantly alleviating the issues of irrelevant responses and hallucinations. In the analysis of financial statements, we have designed the iterative analysis process that allows multiple LLMs to reason and analyze step by step, much like humans do. This process enables the models to derive insights into a company's fundamentals and output their confidence levels.

The contribution of our paper mainly lines in the three fields: 
(1) We collect and open source a multimodal small-scare financial dataset to facilitate further research, which includes news articles, historical stock prices, and financial statement reports. Due to the fact that previous studies have mostly used large-scale datasets and overlooked the position of small-scare individual investors in the market, our dataset is obtained through public channels, reflecting the information that the general public, including retail investors, can strive to access.
(2) We propose a novel Human-Agent collaboration framework, which inspired by a MoE approach, thus can be able to handle multimodal financial data analysis and human interaction. To mitigate hallucinations in LLMs when processing current news, we implement adaptive RAG, ensuring accurate and reliable outputs.
(3) We conduct a comprehensive evaluation and comparison of FinArnea against multiple baseline models across various typical stocks. By designing an input module that captures investors' risk preferences, we are able to identify optimal timing opportunities. This innovation transforms our framework into a Human-Agent collaboration framework, enhancing its practicality and user engagement. Additionally, we discuss the different experiment performance of our approach between the A-share and U.S. stock market. 

The remainder of this paper is organized as follows: we first
introduce the related work in the field of financial market prediction from the perspective of data feature development (§~\ref{Section 2}). Then, we represent FinArena's framework and its technical details of each part (§~\ref{Section 3}).
We display our raw dataset's overview and pre-processing analysis before our experiment (§~\ref{Section 4}). 
Furthermore, we record the results in tabular form of our two experiments with different focuses, accompanied by our technical analysis (§~\ref{Section 5}).
After that, we discuss and reflect the different results we get in the two market (§~\ref{Section 6}). 
Lastly, we summarize the results and discussion, and conclude further potential research directions (§~\ref{Section 7}).

\section{Related Work}
This paper investigates the capabilities of a multimodal, multi-agent and Human-Agent collaboration architecture in financial analysis and forecasting tasks. In this regard, this section reviews the characteristics of the data in the financial forecasting process (§\ref{2.1}), the application of LLM in the field of financial analysis (§\ref{2.2}), and the defects of multi-agent architecture lacking human interaction (§\ref{2.3}).
\label{Section 2}
\subsection{Financial Modeling based on Historical Stock Price}
\label{2.1}
The financial market, with its inherent randomness and volatility, defies a precise description through deterministic models, necessitating the development of models that can guide investment decisions amid uncertainty \cite{Sargent2001RobustCA}. A central focus in financial research is to capture the unique characteristics of financial data to enhance  accuracy \cite{SEZER2020106181}. Historical stock price data, the most accessible, has been the primary source for time series analysis and prediction. The advent of models such as ARIMA\cite{Box1970DistributionOR,Box1978TimeSA} has shifted financial analysis from empirical decision-making to statistically grounded predictions. However, the non-stationarity of financial time series and the tendency for high-order differencing to result in white noise have limited the effectiveness of ARIMA and other similar models. Although the Generalized Autoregressive Conditional Heteroskedasticity Model (GARCH) \cite{BOLLERSLEV1986307}  is specifically designed to capture and simulate volatility clusters in financial and economic time series data, it struggle to capture shocks and extreme values, and the problems of difficult calculations and model complexity make the admission threshold extremely high \cite{Gomes2014}, which is very unfavorable for ordinary investors. 

Machine learning methods, particularly the CNN network \cite{726791} and the LSTM network\cite{6795963}, have marked a significant advance in the prediction of financial time series. These methods can capture more temporal features for better financial time series analysis \cite{10577333}. Despite their improved performance, these methods have the prominent issue of the lag effect. In an attempt to address the limitation, hybrid models have been proposed that combine traditional and machine learning approaches.  \cite{SAADAOUI2024123539} employs an ARIMA-LSTM hybrid model with non-Gaussian distributions to improve financial forecasting. WT-ARIMA-LSTM \cite{ZHANG2024102022} adds a wavelet transform for time series decomposition to extract data features at different time scales, uses ARIMA to smooth the data features and extract linear characteristics, and chooses LSTM to extract non-linear data features that contain noise. Although innovative, these kinds of hybrid models still rely on single historical stock price data, failing to incorporate a broader range of financial market characteristics and thus not achieving a breakthrough in predictive performance.

The reliance on a single data feature, historical stock prices, in these models highlights a significant gap in the current financial modeling landscape \cite{10825449}. The pursuit of more accurate and robust predictive models continues, with the understanding that integrating a wider array of financial data features and addressing the issues of latency and interpretability are crucial for the development of models that can truly meet the demands of the financial market.

\subsection{LLM Agents for Financial Market Analysis}
\label{2.2}
Traditional models of stock price prediction are based on the assumption that the price of a stock on any given day is dependent on the prices of the preceding days \cite{HAREL2021249}. This assertion is increasingly recognized as an oversimplification that does not hold true in the complex dynamics of financial markets \cite{CHEN2024107184}. Consequently, reliance on historical stock price data alone for modeling purposes is deemed insufficient as it fails to capture the broader spectrum of information that influences market behavior.

Beyond structured data of stock prices, financial markets are replete with unstructured data that traditional models have struggled to incorporate \cite{FLOOD2016180,MURINDE2022102103}. Conversely, LLMs excel well at extracting unstructured information, so that they can be naturally utilized for financial unstructured information processing and forecasting. Classified by parameter quantity, it can be divided into two types. The ultra-large models, such as OpenAI's advanced model GPT-3 or GPT-4 \cite{DBLP:journals/corr/abs-2005-14165,openai2024gpt4technicalreport}, are commonly applied to the analysis of financial texts, and their large number of  parameters help to achieve a high level of accuracy in their results. Recent research evaluates ChatGPT's potential to act as a financial advisor \cite{XiaoyangLi2024}, resulting that it has significantly corrected the optimistic bias from human analysts in the financial markets. The other type involves Fine-tuning LLM to optimize their performance in the financial sector. Earlier on appeared the FinBERT series, which are fine-tuned models of BERT. Subsequently in 2023, BLOOM introduced BloombergGPT and Chinese-adapted model Xuanyuan 2.0 \cite{xuanyuan}, Google introduced BBT-Fin \cite{bbtfin}, and a large number of excellent fine-tuned models have emerged from the open-source LLaMA Series \cite{PIXIU,Yang2023InvestLMAL,Touvron2023Llama2O}. This marks the beginning of extensive exploration of LLM's ability to utilized the unstructured data in financial market processing.

Furthermore, these advances in LLMs have revolutionized the way financial markets are analyzed, providing a more nuanced understanding of market dynamics and enhancing predictive accuracy. As a result, the integration of LLMs into financial modeling is set to become a cornerstone in the evolution of methodologies for financial market analysis.

\subsection{Discussion on the Existing Mutil-Agent LLM Systems}
\label{2.3}
With the introduction of an increasing variety of financial data, such as information in the form of text, audio, images, table charts and so on, the effect of using one general LLM for processing and analysis is far from satisfactory \cite{QI2023103510}. This is usually due to the deficient and pointless pre-training data, resulting LLM's ability to capture certain specific elements of these multimodal information is insufficient. Consequently, having specific ``experts'' to take advantage of each specific data has become a new research direction. 

Some systems have assembled multiple LLMs that each specialize in different content. For instance, RiskLabs \cite{Cao2024RiskLabsPF} combines textual and vocal information from Earnings Conference Calls (ECCs), market-related time series and corresponding news data. Framework SEP\cite{10.1145/3589334.3645611} deploys two LLM units respectively for multi-data processing and reflection . FinAgent \cite{Zhang2024AMF} deploys LLMs to process multi-data such as text and images and using a reflection module to analyze the prediction results. These explorations and attempts effectively enhancing the investment performance. 

Nonetheless, there are still imperfections in the current research. First of all, previous LLM frameworks using multimodal data have performed well in financial analysis, but the scare of data they utilized is too large and the cost is too high, which does not show the advantages for retail investors, general investors and low-cost suitors. Secondly, previous LLM frameworks rely heavily on the knowledge of model pre-training, and are prone to hallucinations, especially in understanding news text data. Furthermore, most of the previous researches were based on the framework of machine, they ignore that LLM can not bear any investors' any losses \cite{KLING2023106489}, leading to ignorance of human factors, especially the attitude of risk preference in the decision-making process. This will lead to the monotonous investment decisions of the past methods, which are not in line with real life. 

Therefore, this study proposes a Human-Agent collaboration framework for stock prediction. Specifically, this framework solves the above three problems. First, we deploy multi-LLM agents to analyze multimodel data. Secondly, an adaptive RAG method is introduced to handle hallucinations in news text analysis. Thirdly, on the basis of multi-LLM of multimodal data, the input module of investors' risk preference is further designed. Through the interaction between investors and LLM, FinArena can make investment decisions in line with investors' personal investment preferences, reaching collaboration between advanced AI agents and insightful humans.

\section{Framework}
\label{Section 3}
\subsection{Framework Overview}
\begin{figure}[h]
\centering
\includegraphics[width=1\textwidth]{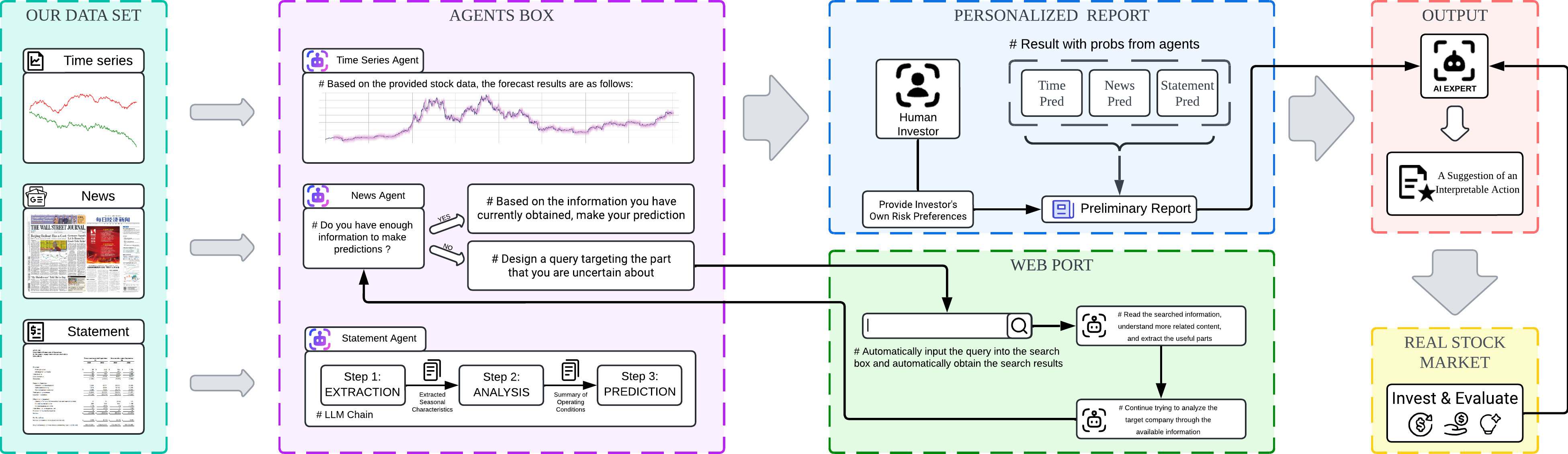}
\caption{The Framework of FinArena}
\label{fig1}
\end{figure}

The workflow of FinArena is delineated in \autoref{fig1}, which encompasses structured and sophisticated system. In a nutshell, FinArena is bifurcated into two different yet interrelated parts: a trio of specialized agents, each excels well at a unique type of data, and one analytical unit tasked with the responsibility of personalized decision-making.

The trio of agents are distinguished by their specialized capabilities in managing diverse data types, ensuring that each category of information is processed with high precision and efficiency. Time Series Agent is adept at processing stock time series data and providing a prediction for future stock prices based on historical information. News Agent specializes in summarizing and extracting insights from news articles, and it is permitted to access to online resources for additional information. Statement Agent, equipped with iterative reasoning capabilities, is specifically tailored for the analysis of structured metrics and indicators within financial statements. The principal task of the analytical unit is to synthesize the outcomes derived from three distinct modalities of data. To formulate personalized investment recommendations, the AI expert utilizes the preliminary report from trio agents and aligns with the individual risk preference of the investors, thus giving a suggestion of an investment action - buy or sell. 

\subsection{Generative Time Series Agent}
Traditional approaches for stock price forecasting have predominantly relied on mathematical models (e.g., ARIMA \cite{Box1970DistributionOR}, GARCH \cite{BOLLERSLEV1986307}) or machine learning algorithms (e.g., XGBoost \cite{Chen2016XGBoostAS}, LSTM \cite{6795963}). These methods, though effective, are fundamentally constrained by the need to construct intricate models and subsequently tailor specific implementation strategies for each model, thereby limiting their applicability across diverse scenarios. In contrast, FinArena employs specialized LLM to address time series analysis task. Time Series Agent fully leverages the extensive corpus of time-series data incorporated during pre-training. Furthermore, it is streamlined and user-friendly, for investor merely needs to send a stock time series dataset as input. Then the agent immediately leverages the vast dataset from its pre-training to forecast the future stock price time series for a predefined period.

The foundation of the forecasting model can be represented as a function $\mathcal{F}_{\theta}: \mathbf{X} \mapsto \mathbf{P}$, which maximizes the conditional probability $\mathcal{P}$, and $\theta$ is a set of all the parameters in the model:
\begin{equation}
    \max\mathcal{P}\left(\mathbf{P}\mid \mathbf{X}\right)=
    \mathcal{F}\left(\mathbf{X};{\theta}\right)
\end{equation}
Where $\mathbf{X}$ is the time-cumulative feature space that includes the historical prices co-variates series $\{(\mathrm{p},\mathrm{c})\}_{0}^{t}$:
\begin{equation}
    \mathbf{X}=
    \left\{\int_0^t \left(\mathrm{p}(s),\mathrm{c}(s) \right) \mathrm{d}s\right\}
\end{equation}
While $\mathbf{P}$ is the forecasting space that includes future prices $\{\mathrm{p}\}_{t+1}^{t+h}$, and the result $\mathbf{P}$ can be regarded as solving the following differential equation as \autoref{equation 3}. $\mathcal{G}\left(\mathbf{P},\mathbf{X},t ; \theta\right)$ is a function dependent on $\mathbf{P}$, feature space $\mathbf{X}$ and time $t$, indicating the rate of change of prediction space over time.
\begin{equation}
\label{equation 3}
    \frac{\mathrm{d}\mathbf{P}}{\mathrm{d}t}=
    \mathcal{G}\left(\mathbf{P},\mathbf{X},t ; \theta\right)
\end{equation}
Our Time Series Agent initially generates the forecasting space $\mathbf{P}$ based on \autoref{equation 3} and represents as \autoref{equation 4} . After that, this agent uses function $\mathcal{H}\left(\cdot \right)$ to transform $\mathbf{P}$ into the 0-1 movement trend as the output time series $\mathbf{O}_{time}$, where $\mathbf{O}_{time}^{k-1}$ refers to the forecast of the stock trend at time $k$:
\begin{equation}
\label{equation 4}
    \mathbf{P}=
    \mathrm{\textit{TimeAgent}}_1\left(\mathbf{X}\right)
\end{equation}
\begin{equation}
    \mathbf{O}_{time}=
    \mathrm{\textit{TimeAgent}}_2\left(\mathbf{P}\right)=
    \left \{\mathcal{H}\left(\mathrm{d}\mathrm{p}\right)\right\}_t^{t+h-1}
\end{equation}

\subsection{Uncertainty-Driven Adaptive News Agent} 
\begin{figure}[h]
\centering
\includegraphics[width=0.85\textwidth]{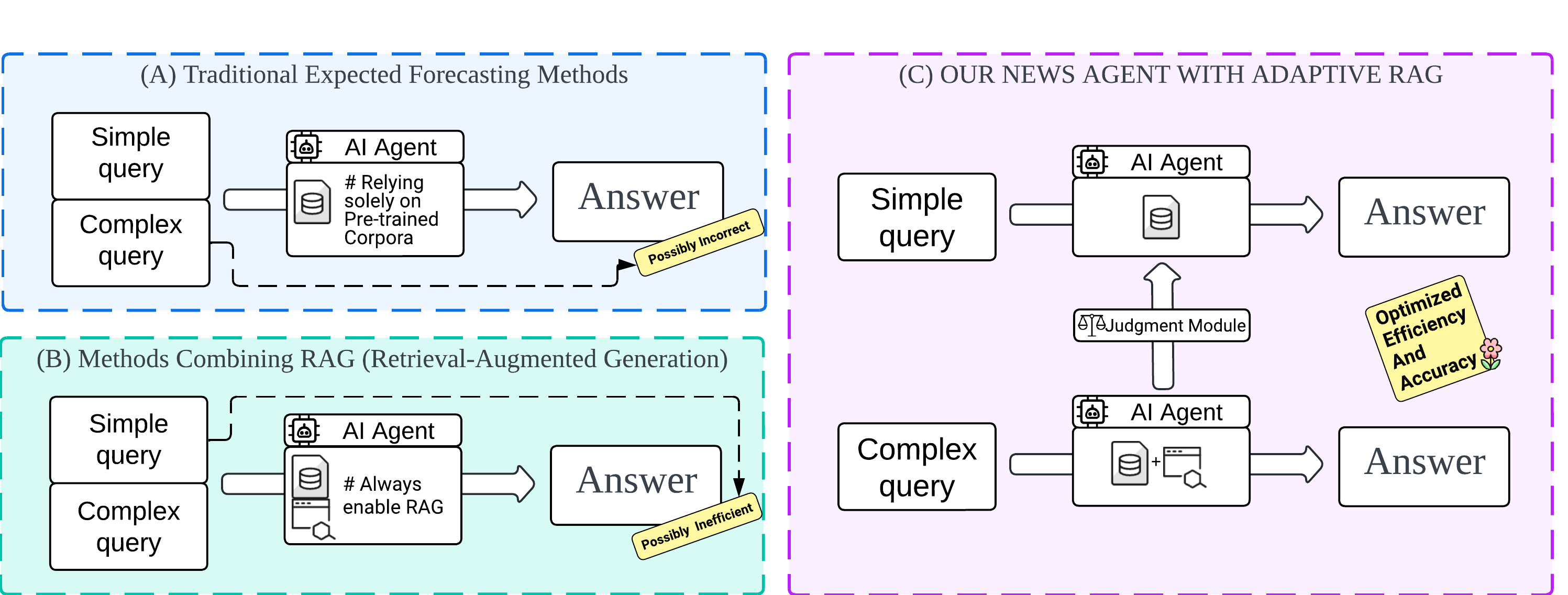}
\caption{The Comparison of Different RAG Application Strategy}
\label{fig2}
\end{figure}

The demand of news article analysis in financial investment decision-making is increasing, yet the processing of long-textural news remains a challenging task. Despite being a powerful and efficient tool for text processing, LLM is not omniscient and may encounter unfamiliar words in news articles that are not contained in their pre-training. RAG is of great help in improving accuracy, efficiency, and personalization \cite{shi2024eragentenhancingretrievalaugmentedlanguage}. On this basis, we design an adaptive RAG approach that enables LLM to perform uncertainty-driven adaptive information retrieval. Our News Agent leverages the strengths of LLM and adapts to dynamic and evolving news content. The comparison is shown in \autoref{fig2}.

Traditional information retrieval and forecasting methods, as depicted in Figure 2(A), depend on static pre-trained corpora, which may not encompass the dynamic vocabulary found in contemporary news. Methods always integrating RAG in Figure 2(B), may lead to an over-reliance on external information sources, thereby reducing the efficiency of handling simple problems that can been directly addressed by their own pre-trained corpora, especially for low-cost-seeking investors, frequent invocation of search engine API is a big expense. News Agent with an uncertainty-driven adaptive RAG introduces a Judgment Module $\left(\mathcal{J}\right)$. It can judge whether the complex query can be directly resolved under the conditions of pre-trained corpora, effectively leveraging LLM while enhancing the accuracy of predictions. The model for news analysis can be represented as \autoref{equation 6}:
\begin{equation}
\label{equation 6}
    \mathbf{S}=\mathrm{\textit{NewsAgent}}_1\left(\mathbf{N}, \mathbf{I}\right)
\end{equation}
Where $\mathbf{N}$ represents the space that encompasses the news article arranged in time $t$ order and item $k$ order $\{n_k\}_{1}^{t}$. Assume the Judge Module ($\mathcal{J}$) is a stochastic process adapted to information filtration $\{\mathcal{L}_t\}_{t \geq 0}$ and is a martingale, meaning that when $\mathcal{J}$ deems the pre-training is sufficient to solve the current task, it returns 0, while it deems insufficient, it returns 1. $\mathcal{J}_{t}$ denotes the value of $\mathcal{J}$ at time $t$, and $\mathbf{I}_k$ is generated as follows:
\[
\mathbf{I}_k =
\begin{cases} 
\emptyset & \text{if } \mathcal{J}_{k} = 0, \\
\{\mathrm{i}_k\}_1^n & \text{if } \mathcal{J}_{k} = 1.
\end{cases}
\]
Here, $\mathbf{I}_k$ is the information set generated at time $k$, depending on the value of $\mathcal{J}_{k}$. $\{\mathrm{i}_k\}_{1}^n$ is the information set from item 1 to item $n$. $\mathbf{S}$ represents as $\{s_k\}_{1}^{t}$ is the summary, analysis and prediction from News Agent, and the output series $\mathbf{O}_{news}$ by the agent is a trend prediction for future stock price movements.
\begin{equation}
    \mathbf{O}_{news}=
\mathrm{\textit{NewsAgent}}_2\left(\mathbf{S}\right)=\left\{\mathcal{H}\left(\mathbf{S}\right)\right\}_{t+1}^{t+h}
\end{equation}

\subsection{Iterative Reasoning Agent for Financial Statement Analysis} 
When analyzing financial statements, human-like judgment enhances the interpretability, transparency and objectivity in the operations of LLMs. In FinArena, we devise an Iterative Reasoning Agent as our Statement Agent. It comprises three distinct steps, and each step's output feeds into the subsequent one, creating a chain of logical progression.

The first step involves an LLM tasked with identifying the seasonal patterns in a company's operations as reflected across four financial statements over a one-year period. This initial analysis is crucial for understanding the cyclical nature of the company's financial health. Next, the seasonal patterns extracted in the first step, along with the original financial statements, are submitted together to the second LLM. It can provide a comprehensive analysis of the company's annual operational conditions at various stages. In the final step, LLM predicts the overall potential rise and fall of the company's stock and assigns a token possibility to reflect its confidence level based on the operational analysis from the previous step. The model for financial report analysis can be represented as:
\begin{equation}
    \mathrm{A}=
    \sum_{Q1\sim FY}\max_{season}\mathrm{\textit{StatementAgent}}_1\left(\mathbf{F}\right)
\end{equation}
\begin{equation}
\mathrm{R}=
\max_\mathcal{P}\left(\mathrm{\textit{StatementAgent}}_2\left(\mathbf{F},\mathrm{A}\right)\right)=
\max \mathcal{P}(\mathrm{R \mid \left(\mathbf{F},\mathrm{A}\right)})
\end{equation}
\begin{equation}
    \mathbf{O}_{statement}=
    \mathcal{H}\left(\max_\mathcal{P}\mathrm{\textit{StatementAgent}}_3\left(\mathbf{F},\mathrm{A},\mathrm{R}\right)\right)
\end{equation}
Where $\mathbf{F}$ represents the four financial statements over a one-year period $\{\mathrm{f}\}_{\mathrm{Q1}}^{\mathrm{FY}}$. $\mathrm{A}$ and $\mathrm{R}$ are intermediate processes in the LLM chain, representing the seasonal characteristic in the statement and LLM's reflection of previous answers. They will be iterated along with the financial statements to the next step. 

The output $ \mathbf{O}_{statement}$ from $\mathrm{\textit{StatementAgent}}_3$ consists of two fields: the long-term trend of the company's stock price change in each season and the LLM's confidence level for its result. Compared to conventional Chain of Thought (CoT) fine-tuning, our Statement Agent fully leverages the financial statement and integrates its own understanding of the financial data each time. Additionally, investor can invoke the Statement Agent of any intermediate process results, thus greatly enhancing its interpretability.

\subsection{Information Aggregation System and Human-Agent Collaboration}
The complexity of financial investment decisions indicates that no single individual can independently complete the entire process of analysis and decision-making anymore. Research \cite{Avkiran2012, BAYRAK2021113490} also mention that effective teamwork is crucial for large-scale problems that exceed the professional capabilities and cognitive abilities of a single decision-maker. In contemporary financial investment landscape, decisions are typically tailored in the form of group collaborations. For instance, the market analyst might focus on market trends and market sentiment, the risk manager might focus on the potential threats, and strategic analysts might focus on the macro and long-term development of the company. We are inspired and migrate this collaboration to multi-agent LLMs, employing selective LLMs which excel at different tasks and aggregating their analysis. For instance, models like TimeGPT \cite{timegpt} or TimesFM \cite{das2024decoderonlyfoundationmodeltimeseries} are suited for temporal data, while models like LLaMA or GPT series are suited for text. This aggregation system realizes the collaboration between multiple agents, and can perform their duties like an investment team to complete more complex investment decision-making tasks.

However, while the collaboration among these specialized agents is essential, it is only part of the solution. In fact, the role of human expertise cannot be overstated. Human involvement in investment decision-making brings a level of nuance, intuition, and adaptability that multi-agent system simply cannot replicate. This is where the concept of Human-Agent collaboration comes into the framework. In FinArena, we design a system that not only facilitates collaboration among different LLM agents but also prioritizes the integration of human expertise. Human investors are not merely passive observers, but also active participants in the decision-making process: they can monitor the analysis results of each agent, review the probability of each response token, and most importantly, interact with the AI Expert. Report Agent provide a port for human-agent interaction, which is set in two places: before giving a rise and fall forecast and before giving investment advice. Investors have the flexibility to interact their risk preferences with LLM agents in either natural language or specific risk metrics, such as ``conservative'' or \textit{Sharpe Ratio}, thus achieve a Human-Agent collaboration for financial investment.

To further illustrate our approach, we underpin our model's capabilities as a function $\mathcal{T}:\left(\mathbf{O},\mathbf{R}\right) \mapsto \mathbf{A}$, that:
\begin{equation}
    \{\mathbf{A} \mid \left(\mathbf{O},\mathbf{R}\right)\}=\mathcal{T}\left(\mathbf{O},\mathbf{R}\right)
\end{equation}
\begin{equation}
    \mathbf{R}=\mathrm{\textit{ReportAgent}}_1\left(\mathrm{RiskPreference}\right)
\end{equation}
$\mathbf{R}$ represents the risk preference from the investor and summarized by LLM, without which LLM can not provide a prediction. While $\mathbf{A}$ can be calculated by :
\begin{equation}
    \mathbf{A}= \mathcal{K}_t\circ \mathrm{\textit{ReportAgent}}_2\left(\mathbf{O},\mathbf{R}\right)
\end{equation}
Where $\mathbf{O}$ represents the output space $\left\{\mathbf{O}_{time},\mathbf{O}_{news},\mathbf{O}_{statement}\right\}$, and $\mathcal{K}_t$ is a state transition correction that adjusts LLM agents' understanding of human feedback. It can be described by the following equation:
\begin{equation}
    \mathcal{K}_{t+1}=\textit{feedback}_t\cdot\mathcal{K}_t\left(\mathbf{R}_{t+1}\right)
\end{equation}

In summary, FinArena highlights the significance of Human-Agent collaboration in the realm of financial investment decision-making in the Report Agent segment. By integrating the strengths of specialized LLM agents and human expertise, our aim is to achieve a more comprehensive and nuanced approach to the solution of complex investment problems. The human-in-the-loop mechanism ensures that investors' unique insights, intuition, and risk preferences are fully incorporated into the decision-making process, rather than being overshadowed by purely algorithmic solutions.

\section{Dataset}
\label{Section 4}
Corporate information often abounds a mixture of structured and unstructured data. The challenge lies in discerning the subset data which is truly available. Financial news in current researches is commonly marked by two predominant ways: utilizing existing publicly datasets, or developing private bespoke datasets. For our investigation, we choose the latter approach to address the gaps inherent in publicly available datasets, which suffer from a lack of relevance, outdated information, or an insufficient length of content: (1) These datasets often include an array of other news topics (e.g., sports, fashion, politics...), which probably dilute the focus on financial factors when processing them simultaneously. (2) The timeliness of publicly data may have already been factored into the pre-training of LLMs, thereby potentially inflating the perceived performance. (3) Many datasets only provide headlines, neglecting the context of the news articles. 

By constructing our own dataset, we aim to rectify these shortcomings, ensuring that our data is not only current and relevant but also can capture the full spectrum of financial news and its nuanced impact on market dynamics. Additionally, researches utilizing publicly available large-scare datasets for testing demonstrate the usability for large investment companies, however, obtaining these vast data from public market is not feasible for any low-cost or small-scale investors. Therefore, our dataset also considers that our dataset scare is not too large or too costly, in order to reflect the effectiveness of FinArena for small-scare or retail investors.

To comprehensively compare and evaluate the performance of FinArena in different market conditions, we carry out our experiments in both the A-share and U.S. stock markets. And to avoid an overestimation of the analytical performance caused by data in the LLM's pre-training datasets as much as possible, we choose the data spanning from January 1st, 2023 to March 30th, 2024. We selected companies based on several dimensions that included company influence, transparency of information, and the availability of comprehensive data, and the chosen companies are those that not only have significant influence within their respective industries but also maintain a high level of information disclosure. In each stock market, we respectively select five companies that best meet our criteria. These companies are listed in \autoref{Table1}, which provides an overview of the data type and sample volume. We have shared raw dataset on Hugging Face to encourage further research \footnote{https://huggingface.co/datasets/Illogicaler/FinArena-low-cost-dataset}.

\begin{table}[htbp]
    \centering
    \begin{tabular}{|c|cccccc|}
    \hline
        \diagbox{\textbf{Company}}{\textbf{Type}}  & Market & Code & Stock & News & k-means & Statements   \\
    \hline
        Amazon & U.S. &AMZN &311 &204& 147& $\surd$ \\
        Google & U.S. &GOOG & 311& 312&205& $\surd$\\
        Microsoft & U.S. &MSFT &311 & 293&197& $\surd$\\
        Nvidia & U.S. &NVDA &311 &193&119 & $\surd$\\
        Tesla & U.S.&TSLA &311 & 289&187& $\surd$\\
        BYD & A-share & 002594 & 300 &203 &142 &$\surd$ \\
        CATL &A-share &300750 &300 & 148&98& $\surd$\\
        CMCC& A-share& 600941&300 & 302&161&$\surd$ \\
        Loongson  & A-share & 688047 &300 &188 &116&$\surd$ \\
        MOUTAI& A-share & 600519  &300 &163 &112& $\surd$\\
    \hline
    \end{tabular}
    \caption{\textbf{Dataset Overview. }The k-means column represents the number of after-preprocessed news article items.}
    \label{Table1}
\end{table}

    \paragraph{Stock Price Data} Our stock price data focuses on indicators that are fundamental to reflect the stock's market performance, and these indicators are also the most basic data that the majority of investors can easily leverage. We call \textit{Tsanghi} API and obtain the opening price, closing price, and trading volume as our indicators, for their accessibility and their proven relevance in financial analysis. The opening price provides a snapshot for the start of the trading day, while the closing price conversely reflects the ultimate attitude. These two indicators offer a clear view of the stock's performance over the course of a single day. Besides, trading volume is regarded as a measure of the liquidity and interest of the stock. 
    
    \paragraph{Financial News} Financial news plays a vital role in shaping market sentiment and influencing investment decisions. The immediacy and accuracy of financial news are crucial for investors to make informed decisions in a dynamic market environment. Public datasets often come with the lack of specificity, recency, and thematic focus. Our approach addresses these limitations by constructing a bespoke dataset that the data is not only more current but also directly applicable to our study, thereby enhancing the relevance and impact of our findings. The selection of authoritative and legally permissible sources for web crawling is essential. For A-share companies, we systematically crawl the National Business Daily website to capture all news articles that include company's specific keywords. Similarly, for U.S. companies, we employ the same methodology on Business Today website. The news data consists of three fields: ``Title'' ``Date'' and ``Text''. We standard the time format, remove irrelevant text such as journalists' or editors' names, and use k-means clustering to select the most representative news article within the same date range. The key steps of our data pre-processing will be presented later.
    
    \paragraph{Financial Statements} Financial statements serve as a comprehensive reflection of a company's past and present operational conditions, which can be utilized for analyzing and forecasting the company's future. However, the full financial reports provided by companies can be extremely extensive when inputting into any LLM, posing challenges in effectively extracting and utilizing pertinent information. We also invoke \textit{Tsanghi} API and select the key financial indicates from the balance sheet, cash flow statement and income statement. Subsequently we combine them as a single input file. The approach of only selecting tabular part and indicators is supported by existing research. When financial tabular data are already available, the incremental value of textual disclosure in financial reports is minimal \cite{ZHAO2023102770}.

\section{Experiments}
\label{Section 5}
In this section, we conduct a thorough evaluation of FinArena framework using the collected dataset, focusing on two key tasks: predicting stock movements and simulating stock trading. Our investigation addresses the following research questions:
\begin{itemize}[itemsep=1pt, topsep=1pt, parsep=1pt, partopsep=2pt, leftmargin=2em]
    \item \textbf{RQ1}: How does FinArena perform in predicting stock movements compared to traditional benchmarks?
    \item \textbf{RQ2}: Is the use of unstructured data, particularly news data, more advantageous when applied extensively?
    \item \textbf{RQ3}: How do individual market expectations affect the decision-making accuracy of AI Experts in the early stages of the task?
    \item \textbf{RQ4}: How can FinArena be effectively utilized, and does an individual's risk preference influence its effectiveness in trading simulations?
\end{itemize}
This section presents a systematic assessment of the FinArena framework, aiming to evaluate its performance in predicting stock movements and simulating trading scenarios. Additionally, we explore how individual preferences and risk appetites impact the framework's results.

\subsection{Experimental Setup}
In our experiments, we utilize the cost-effective DeepSeek-v2 \cite{deepseekai2024deepseekv2strongeconomicalefficient} model to create the News Agent and the Statement Agent, which are adept at handling unstructured data. For processing historical stock price data, we employed TimeGPT \cite{timegpt}, a model designed for large-scale time series forecasting, to develop the Stock Agent. The final AI Expert was constructed using the gpt-4o-mini model. To further investigate \textbf{RQ2}, we also created unstructured data agents using LLAMA-3-70B model \cite{dubey2024llama3herdmodels}, which excels in processing English text, and Kimi (moonshot-v1-32k) \cite{Moonshot}, which is effective for Chinese text.

In Stock Movement Prediction, FinArena operates in prediction mode, where the AI Expert generates a binary forecast indicating whether a stock is expected to rise or fall the following trading day. We evaluated methodologies using binary classification metrics, including accuracy (\textbf{Acc}) and F1-Score (\textbf{F1}).

In stock Trading Simulation, FinArena translates trend predictions into actionable recommendations to buy, sell, or hold stocks, which are then tested through actual investments based on these recommendations. We considered four distinct risk profiles: conservative, moderately conservative, moderately aggressive, and aggressive. Each profile differs in how funds are allocated during transactions.

\begin{itemize}[itemsep=1pt, topsep=1pt, parsep=1pt, partopsep=2pt, leftmargin=2em]
    \item \textbf{[Cons.]} Conservative investors allocate 50\% of their idle funds for each trade and withdraw 100\% of their funds when risk is anticipated. 
    \item \textbf{[M.Cons.]} Moderately conservative investors use 70\% of their idle funds per trade and also withdraw 100\% upon risk prediction. 
    \item \textbf{[M.Agg.]} Moderately aggressive investors invest 100\% of their idle funds each time and withdraw 50\% when risk is predicted. 
    \item \textbf{[Agg.]} Aggressive investors similarly invest 100\% of their idle funds, but each time withdraw only 30\% upon prediction of the risk. 
\end{itemize}

We evaluated the performance of FinArena and the baseline models using metrics such as Annualized Return (\textbf{AR}), Sharpe Ratio (\textbf{SR}) and Maximum Drawdown Ratio (\textbf{MD}).

\subsection{Baselines}
In the task of predicting stock movements, we compare FinArena with several benchmark models. These benchmarks include traditional approaches ARIMA and LSTM networks, as well as the Time Series LLM, TimeGPT.

For the ARIMA model, we used the AIC and BIC criteria to determine the appropriate differencing order as 2 during experimental preparation. Furthermore, we employed grid search to identify suitable lag orders, attempting to capture as many nonlinear characteristics in the stock price data as possible. There are indications that using general models such as ARIMA(1,2,1) would further degrade the predictive capability for stock price time series. We also trained a small LSTM model as one of our baseline models. It consists of 4 layers, trained over 200 epochs with a batch size of 32. We employ a single TimeGPT as another baseline, it automatically sets the opening price, highest price, and lowest price as exogenous co-variates to predict the changes in the closing price. 

In the context of stock trading simulation, our comparative benchmarks encompass random strategies and specific trading methodologies. The Buy-on-Rising-Streak and Sell-on-Falling-Streak (BRSF) strategies are based on short-term market trends, where investors purchase stocks during periods of consecutive price increases and sell during periods of consecutive price declines. In addition, we evaluate strategies based on the ARIMA and LSTM models. To demonstrate the interesting performance of news information analysis, we supplemented the results of experiments that  relied solely on news data for predicting stock price movements, serving as another baseline for our study after completing the experiments. All methodologies have been rigorously assessed to ensure a comprehensive performance comparison in our datasets.

\subsection{Metrics}
\subsubsection{Stock Movement Prediction}
\begin{figure}[htbp]
\centering
\includegraphics[width=0.9\textwidth]{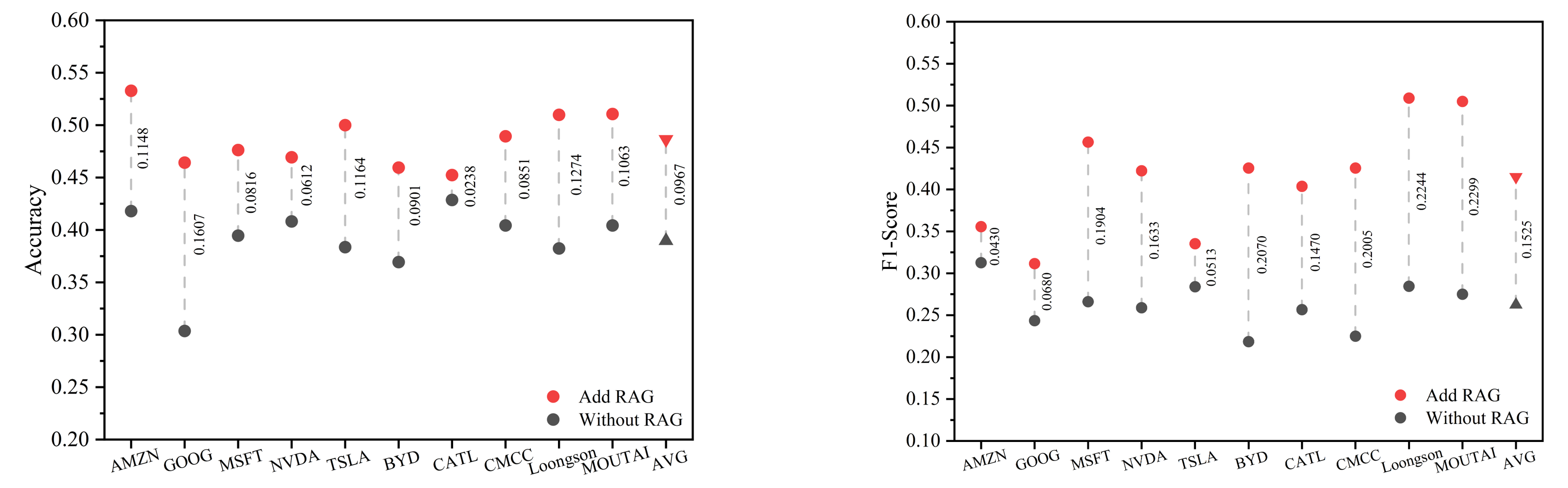}
\caption{\textbf{The Results of Ablation Experiment.} Using RAG or not are marked with \textcolor{orange}{\textbf{orange}} and \textcolor{gray}{\textbf{gray}} respectively, and the dotted line is used to connect the two and display the difference between them. The triangle's dots represent the average performance of all companies.}
\label{fig4}
\end{figure}

In our experimental framework, all baseline models were either implemented by us or derived from existing open-source implementations. First, we tested Accuracy and F1-Score of various methods, and the results were shown in \autoref{table2} \& \autoref{table3}. Then we used the news data set to conduct an ablation experiment of our adaptive-RAG method against the non-RAG method, and the results are shown in \autoref{fig4}.

\begin{table}[htbp]
\centering
\resizebox{\linewidth}{!}{
\begin{tabular}{|l|ccccc|ccc|}
\hline
\diagbox{\textbf{Company}}{\textbf{Methods}} & \textbf{ARIMA} & \textbf{LSTM} & \textbf{TimeGPT}  & \textbf{FinArena.D} & \textbf{FinArena.L/K}  & \textbf{Sen.} & \textbf{Insen.} & \textbf{Opt.}\\
\hline
Amazon & 0.4333 & 0.3770 & 0.5000  & 0.5500 & 0.5500  & 0.5333 & 0.5167 & 0.5167 \\
Google & 0.5000 & 0.5410 & 0.6167  & 0.5833 & 0.5500  & 0.5667 & 0.6000 & 0.5500 \\
Microsoft & 0.4500 & 0.5574 & 0.6333  & 0.6000 & 0.6167  & 0.5833 & 0.6000 & 0.6667\\
Nvidia & 0.5833 & 0.4918 & 0.4667  & 0.5167 & 0.5167  & 0.4833 & 0.5000 & 0.5500\\
Tesla & 0.5000 & 0.4590 & 0.6557  & 0.6667 & 0.6667  & 0.6667 & 0.6500 & 0.5667\\
\hline
\textbf{Mean$\uparrow$} & 0.4933 & 0.4852 & 0.5745  & \textcolor{orange}{\textbf{0.5833}}& 0.5800  & 0.5667 & \textcolor{orange}{\textbf{0.5733}}& 0.5700\\
\textbf{Std.$\downarrow$} & 0.0523 & 0.0644 & 0.0762  & \textcolor{orange}{\textbf{0.0506}}& 0.0542  & 0.0606 & 0.0564 & \textcolor{orange}{\textbf{0.0510}}\\
\hline
BYD & 0.4561 & 0.5357 & 0.5263  & 0.5179 & 0.5179  & 0.5000 & 0.5357 & 0.5000 \\
CATL & 0.5439 & 0.4912 & 0.4561  & 0.4643 & 0.4821  & 0.4821 & 0.4821 & 0.5000 \\
CMCC & 0.3684 & 0.5263 & 0.4737  & 0.5000 & 0.4643  & 0.4286 & 0.5000 & 0.4643 \\
Loongson & 0.5614 & 0.5614 & 0.4561  & 0.5357 & 0.5179  & 0.5000 & 0.4821 & 0.4464 \\
MOUTAI& 0.4912 & 0.4386 & 0.5362  & 0.5179 & 0.5179  & 0.5179 & 0.5179 & 0.5000 \\
\hline
\textbf{Mean$\uparrow$} & 0.4842 & \textcolor{orange}{\textbf{0.5106}}& 0.4897  & 0.5072 & 0.5000  & 0.4857 &\textcolor{orange}{\textbf{0.5036}}& 0.4821\\
\textbf{Std.$\downarrow$} & 0.0690 & 0.0425 & 0.0347  & 0.0242 & \textcolor{orange}{\textbf{0.0226}}& 0.0307 & \textcolor{orange}{\textbf{0.0208}}& 0.0226 \\
\hline
\end{tabular}
}
\caption{\textbf{Accuracy of Stock Movement Prediction Task.} Left of the vertical line: model type test. Right of the vertical line: fixed model type with investor attitude changes. Optimal results highlighted in \textcolor{orange}{\textbf{orange}}. Table shows: (1) Prediction accuracy for FinArena and baselines. (2) Test results for replacing NewAgent LLM with specialized language model. (3) Fixed LLM as benchmark in FinArena, with investor understanding test results during prediction phase.}
\label{table2}
\end{table}

\begin{table}[t!]
\centering
\resizebox{\linewidth}{!}{
\begin{tabular}{|l|ccccc|ccc|} 
\hline 
\diagbox{\textbf{Company}}{\textbf{Methods}} & \textbf{ARIMA} & \textbf{LSTM} & \textbf{TimeGPT} & \textbf{FinArena.D} & \textbf{FinArena.L/K} & \textbf{Sen.} & \textbf{Insen.} & \textbf{Opt.}\\
\hline 
Amazon & 0.3773 & 0.3552 & 0.5000 & 0.4128 & 0.4357 & 0.5200 & 0.5133 & 0.3939 \\ 
Google & 0.4792 & 0.5065 & 0.6166 & 0.5823 & 0.5469 & 0.5662 & 0.6000 & 0.5438 \\
Microsoft & 0.4020 & 0.5438 & 0.6296 & 0.5982 & 0.6157 & 0.5804 & 0.5960 & 0.6633 \\ 
Nvidia & 0.4316 & 0.3748 & 0.4661 & 0.5133 & 0.5133 & 0.4820 & 0.4978 & 0.4725 \\ 
Tesla & 0.4857 & 0.4373 & 0.6491 & 0.6633 & 0.6633 & 0.6633 & 0.6452 & 0.5647 \\ 
\hline 
\textbf{Mean$\uparrow$} & 0.4352 & 0.4435 & \textcolor{orange}{\textbf{0.5723}}& 0.5540 & 0.5550 & 0.5624 & \textcolor{orange}{\textbf{0.5704}}& 0.5276 \\ 
\textbf{Std.$\downarrow$} & \textcolor{orange}{\textbf{0.0423}}& 0.0729 & 0.0744 & 0.0852 & 0.0793 & 0.0613 & \textcolor{orange}{\textbf{0.0560}}& 0.0905 \\ 
\hline 
BYD & 0.4242 & 0.5303 & 0.5165 & 0.5165 & 0.5165 & 0.5000 & 0.5351 & 0.4556 \\ 
CATL & 0.4805 & 0.4974 & 0.4531 & 0.4615 & 0.4780 & 0.4376 & 0.4780 & 0.4667 \\ 
CMCC & 0.3700 & 0.5351 & 0.4636 & 0.5000 & 0.4643 & 0.4167 & 0.5000 & 0.4531 \\ 
Loongson & 0.5281 & 0.5709 & 0.4615 & 0.5204 & 0.4985 & 0.3875 & 0.4614 & 0.4376 \\ 
MOUTAI& 0.4974 & 0.4242 & 0.5333 & 0.5165 & 0.5165 & 0.5102 & 0.5165 & 0.4556 \\ 
\hline \textbf{Mean$\uparrow$} & 0.4601 & \textcolor{orange}{\textbf{0.5116}}& 0.4856 & 0.5030 & 0.4947 & 0.4504 & \textcolor{orange}{\textbf{0.4982}}& 0.4537 \\ 
\textbf{Std.$\downarrow$} & 0.0563 & 0.0495 & 0.0327 & 0.0219 & \textcolor{orange}{\textbf{0.0208}}& 0.0475 & 0.0263 & \textcolor{orange}{\textbf{0.0093}}\\ 
\hline 
\end{tabular}
}
\caption{\textbf{F1-Score of Stock Movement Prediction Task.} Left of the vertical line: model type test. Right of the vertical line: fixed model type with investor attitude changes. Optimal results highlighted in \textcolor{orange}{\textbf{orange}}. Table shows: (1) F1-Score for FinArena and baselines. (2) Test results for replacing NewAgent LLM with specialized language model. (3) Fixed LLM as benchmark in FinArena, with investor understanding test results during prediction phase.}
\label{table3}
\end{table}

Here we summarize three main findings: (1) In eight of the ten cases, FinArena's performance can beat the multiple baselines we set, not only achieving a higher average, but also having a smaller variance. This implies that multimodal data analysis methods can outperform single stock price timing analysis, whether one-dimensional or multi-dimensional. This is the answer to \textbf{RQ1} \& \textbf{RQ2}. (2) Compared to DeepSeek-v2, the language-specialized model Kimi or LLaMA 3, exhibits a certain degree of performance decline. This suggests that an excessive focus on the language style and expression used in news reporting may lead to a decrease in the accuracy of analysis, which means that the introduction of human biases during the model analysis phase is inappropriate. (3) When using DeepSeek-v2, pre-informing the model about investors' attitudes towards the market (sensitive, insensitive, or optimistic) also results in a decline in model performance. Among these three attitudes, the insensitive attitude has the smallest impact on model performance. Therefore, our response to \textbf{RQ3} is that it is unwise to introduce investors' market expectations during the FinArena forecasting phase. (4) The addition of RAG significantly improved the performance of the model in terms of accuracy and F1-Score. The improvements are consistent across companies. In A-share companies, our adaptive-RAG has achieved more significant results. This may be due to the higher complexity and confusion of the A-share information.

\subsubsection{Stock Trading Simulation}
To evaluate the performance of FinArena compared with baseline models, we conduct a comprehensive backtesting procedure. This backtesting period was aligned with the test set duration of the dataset, specifically accounting for data loss incurred through ARIMA differencing, covering the timeframe from January 1, 2024, to March 27, 2024. All results have been annualized to facilitate direct comparison. The findings from this analysis are detailed in \autoref{Table4}.

\begin{table}[htbp]
\centering
\resizebox{\linewidth}{!}{
\begin{tabular}{|l|cccccccc|}
\hline
\diagbox{\textbf{Company}}{\textbf{Strategy}} & Random & BRSF & ARIMA & LSTM & Cons. & M.Cons. & M.Agg. & Agg. \\
\hline
\textbf{AR} &&&&&&&&\\
\hline
Amazon & 0.4149& 0.2229 & 0.2775 & 0.0650 & 0.8383 & 0.6685 & 0.9504 & 0.9386 \\
Google &\textcolor{red}{(0.0214)}&0.1188 & 0.1404 & 0.1188 & 0.5896 & 0.4169 & 0.4770 & 0.6049 \\
Microsoft &0.1589 & 0.1268 & 0.2159 & 0.1578 & 0.3833 & 0.2143 & 0.5863 & 0.5383 \\
Nvidia &0.2792 &4.8994 & 2.0266 & 1.2018 & 0.6530 & / & 3.4535 & 2.1140 \\
Tesla &\textcolor{red}{(0.4605)}&\textcolor{red}{(0.2253)}& \textcolor{red}{(0.3958)}& \textcolor{red}{(0.3337)}& 0.3762 & 0.3600 & \textcolor{red}{(0.0270)}& 0.1708 \\
BYD & 0.0858&0.1807 & 0.0089 & 0.2925 & 0.3221 & 0.2473 & 0.4488 & 0.4316 \\
CATL &0.3592 &\textcolor{red}{(0.0799)}& 0.0787 & 0.1372 & \textcolor{red}{(0.0221)}& \textcolor{red}{(0.0044)}& 0.5845 & 0.3508 \\
CMCC &0.1635 &0.0036 & 0.0827 & 0.1086 & 0.1410 & (0.0281) & 0.1853 & 0.1566 \\
Loongson &\textcolor{red}{(0.6407)}&0.1010 & \textcolor{red}{(0.1233)}& 0.2549 & \textcolor{red}{(0.1233)}& \textcolor{red}{(0.1525)}& \textcolor{red}{(0.3194)}& \textcolor{red}{(0.2695)}\\
MOUTAI& 0.0259&\textcolor{red}{(0.0550)}& 0.0015 & \textcolor{red}{(0.1045)}& \textcolor{red}{(0.0046)}& 0.0531 & \textcolor{red}{(0.0681)}& \textcolor{red}{(0.0353)}\\
\hline
\textbf{Mean$\uparrow$} & 0.0365& 0.5293 & 0.2313 & 0.1898 & 0.3153 & 0.1973 & \textcolor{blue}{\textbf{0.6271}} & 0.5001 \\
\hline
\textbf{MD}&&&&&&&&\\
\hline
Amazon &0.0249 &0.0317 & 0.0309 & 0.0308 & 0.0390 & 0.0332 & 0.0395 & 0.0401 \\
Google &0.1208 &0.0876 & 0.0904 & 0.0999 & 0.0336 & 0.0259 & 0.0738 & 0.0545 \\
Microsoft &0.0195 &0.0194 & 0.0228 & 0.0232 & 0.0233 & 0.0118 & 0.0292 & 0.0244 \\
Nvidia &0.0060 &0.0681 & 0.0639 & 0.0379 & 0.0754 & / & 0.0624 & 0.0560 \\
Tesla &0.1672 &0.0681 & 0.1730 & 0.1155 & 0.1062 & 0.0734 & 0.1538 & 0.1508 \\
BYD & 0.0483&0.0463 & 0.0604 & 0.0589 & 0.0440 & 0.0378 & 0.1024 & 0.0852 \\
CATL & 0.0454&0.0376 & 0.0274 & 0.0349 & 0.0724 & 0.0599 & 0.0830 & 0.0743 \\
CMCC &0.0167 &0.0469 & 0.0385 & 0.0321 & 0.0486 & 0.0499 & 0.0391 & 0.0410 \\
Loongson &0.2334 & 0.1717 & 0.2106 & 0.1033 & 0.1583 & 0.1339 & 0.2465 & 0.2236 \\
MOUTAI&0.0260 &0.0219 & 0.0248 & 0.0315 & 0.0527 & 0.0386 & 0.0600 & 0.0580 \\
\hline
\textbf{Mean$\downarrow$} & 0.0708& 0.0599 & 0.0743 & 0.0568 &0.0653 & \textcolor{blue}{\textbf{0.0516}} & 0.0890 & 0.0808 \\
\hline
\textbf{SR}&&&&&&&& \\
\hline
Amazon & 4.1954&1.6674 & 1.9697 & 0.3366 & 3.4674 & 3.2615 & 3.7878 & 3.7113 \\
Google &\textcolor{red}{(0.4469)}&0.4942 & 0.6886 & 0.4885 & 4.7501 & 4.4051 & 2.4280 & 3.3641 \\
Microsoft &2.2204 &1.4570 & 2.7491 & 1.9261 & 4.0097 & 3.7502 & 4.3148 & 4.7709 \\
Nvidia &3.7680 &14.8063 & 5.3308 & 7.2679 & 2.9032 & / & 11.1131 & 8.2388 \\
Tesla & \textcolor{red}{(4.3782)}&\textcolor{red}{(1.8892)}& \textcolor{red}{(1.6351)}& \textcolor{red}{(1.9701)}& 1.2749 & 1.6530 & \textcolor{red}{(0.1457)}& 0.3845 \\
BYD & 0.7797&0.9730 & \textcolor{red}{(0.0153)}& 1.6166 & 1.7966 & 1.5566 & 1.8264 & 1.9005 \\
CATL &2.7907 &\textcolor{red}{(0.9208)}& 0.7309 & 0.8031 & \textcolor{red}{(0.1837)}& \textcolor{red}{(0.1012)}& 1.7377 & 1.2033 \\
CMCC &1.8963 &\textcolor{red}{(0.0483)}& 0.5445 & 0.7571 & 1.0753 & (0.4017) & 1.0987 & 0.9899 \\
Loongson &\textcolor{red}{(4.3782)}& 0.2068 & \textcolor{red}{(0.2962)}& 0.7212 & \textcolor{red}{(0.4236)}& \textcolor{red}{(0.6594)}& \textcolor{red}{(0.8052)}& \textcolor{red}{(0.7543)}\\
MOUTAI&0.2012 &(0.9943) & (0.1132) & \textcolor{red}{(1.7502)}& \textcolor{red}{(0.1175)}& 0.3879 & \textcolor{red}{(0.5388)}& \textcolor{red}{(0.3153)}\\
\hline
\textbf{Mean$\uparrow$} & 0.6671& 1.5752 & 0.9954 & 1.0197 & 1.8553 & 1.5391 & \textcolor{blue}{\textbf{2.4817}} & 2.3494 \\
\hline
\end{tabular}
}
\caption{\textbf{The Performance in Stock Trading Simulation Task.} (1) This table shows the annualized rate of return (AR), maximum withdrawal rate (MD) and Sharpe rate (SR) of investment with an initial capital of 100,000 USD/CNY under different investment strategies. (2) The best of each indicator will be displayed in \textcolor{blue}{\textbf{blue}}. (3) The value in brackets and \textcolor{red}{red} is negative, and the slash (/) indicates that investment cannot be made in this way.}
\label{Table4}
\end{table}

In this experiment, though the benefits brought by various strategies based on FinArena prediction results vary, they generally exceed the baseline. Among the four investor's risk preference, M.Agg achieves the best \textbf{AR} and \textbf{SR} performance, but the worst \textbf{MD} performance. This helps to answer \textbf{QR4}. We believe that different risk preferences can affect the final investment return of FinArena, and the combination of personal risk preferences with FinArena is demonstrated meaningful. Specifically, the absolute size of the final return is still in the hands of investors, and FinArena improves the lower bound of returns by analyzing and integrating various information.

Our dataset simulates the multimodal data that average investors can obtain at a low cost. In this dataset, FinArena still has an advantage in the investment of multiple stocks. This indirectly validates the effectiveness of our Human-Agent collaboration architecture itself. However, in reality, FinArena only achieves the best on average, rather than performing well on all stocks. For example, in Nvidia, which has an abnormally sharp increasing trend, the BRSF ``empirical method'' even outperforms various other methods, achieving the highest \textbf{AR} and \textbf{SR} results. In the A-share market, all FinArena investment methods lead to negative returns, while the baseline methods can achieve positive returns. These poor performances in both markets are attributed to incomplete and inadequate multimodal datasets of individual investors, which may lead to possible research directions in the future, that is, how to make full use of the incomplete datasets to obtain more accurate analysis and predictions as much as possible, so that LLM can also serve small-scale and low-cost investors.

\section{Discussion}
\label{Section 6}
Our findings indicate that FinArena outperforms other methods in predictive performance on the U.S. stock market dataset. By integrating diverse information sources, it achieves both the highest accuracy and the lowest variance, demonstrating the stability of its predictions. This suggests that profit-related insights in the U.S. market are well-distributed across different data types. However, when applied to the A-share market dataset, FinArena's effectiveness is lower.
Meanwhile, although traditional machine learning models (i.e., LSTM) show advantages in predicting certain stocks, their overall performance remains mediocre. 

We speculate that this may stem from underlying differences in market structure or data distribution, as suggested by issues we identified from the early stages of data collection on the A-share market. Although our study primarily selected companies based on market capitalization, we found that many firms had insufficient news coverage on financial websites. Additionally, we observed an abnormally high prevalence of positive news, highlighting a significant degree of information asymmetry in the market. Similar findings have been discussed in previous studies \cite{yuan201649, LI2020101293}.

We hypothesize that the underlying factors may include regulatory constraints on information disclosure imposed by macroeconomic policies. For instance, during periods aimed at stimulating market vitality, restrictions may be placed on the dissemination of negative news. Furthermore, delays in internal information transmission within the A-share market may result in negative news being overshadowed by positive reports. Additionally, the K-means clustering mechanism used in data processing may have amplified this phenomenon, and the adapted RAG method we employed still faces problems such as repeatedly searching for the same external knowledge, the information retrieved by a single Chinese query is limited and insufficient, the issue of higher complexity in Chinese and so on, leading to this phenomenon as well \cite{shi2024enhancingretrievalmanagingretrieval}. Further research is needed to explore these factors in depth.

Despite the challenges related to information disclosure and market efficiency in the A-share market, investment strategy preferences also play a crucial role. When all investors rely on FinArena’s predictions, differences in investment strategies significantly impact final returns. Notably, in both markets, no single investment strategy demonstrates a clear dominant advantage. This finding suggests that, while the A-share market has room for improvement, it would be overly pessimistic to dismiss it as entirely ineffective or unreliable. Instead, these results highlight the complexity and diversity of investment strategies across markets, emphasizing the interplay between market structure, information accessibility, and investor behavior.

\section{Conclusion}
\label{Section 7}
In this study, we propose FinArena, an innovative Human-Agent collaboration framework for financial analysis and forecasting. Specifically, FinArena imitates a MoE approach to employ several specific LLM agents to process multimodal financial data analysis and human interaction. By combining the strengths of multiple LLMs specialized in different types of financial data and human preference, FinArena aims to provide more accurate and reliable predictions for stock movements and support personalized investment decision-making.
Our comprehensive experiments demonstrate that FinArena outperformed several traditional and advanced benchmarks in predicting stock movements, achieving higher accuracy and F1-Scores. The integration of multimodal data, including stock prices, news articles, and financial statements, proved to be more effective than relying on a single source of stock price data. Additionally, the adaptive RAG method employed in the News Agent can reduce the occurrence of hallucinations or irrelevant responses and lower the cost, thus enhancing the model's performance.

In the stock trading simulation task, FinArena shows promising results, with different risk preferences yielding varying returns. The moderately aggressive strategy achieves the best annualized return (AR) and Sharpe ratio (SR), while the conservative strategy provided better risk management. These findings highlight the importance of aligning investment strategies with individual risk preferences and underscore the potential of FinArena to improve investment outcomes for retail investors.

Our analysis also reveals differences in the effectiveness of FinArena across the U.S. and A-share markets. Although FinArena performs exceptionally well in the U.S. market, its performance in the A-share market was less consistent, likely due to issues related to information disclosure and data quality. This suggests that the maturity and transparency of financial markets play a crucial role in the performance of multi-agent LLM investment models, which means retail or small-scale investors need to make reasonable choices in the investment market in order to maximize LLM's information processing capabilities, and the public disclosure resources easily obtained and utilized at present can not always make these small-scare investors "Rake in money without lifting a finger".

In conclusion, FinArena represents a significant advance in the application of LLMs for financial analysis. Its multimodal approach and adaptive mechanisms address key challenges in traditional financial modeling and provide a robust solution for personalized investment decision-making. Future work may focus on further improving the model's adaptability to different market conditions, enhancing data quality in emerging markets, and exploring additional multimodal data sources to enrich the framework's capabilities.

\bibliographystyle{unsrt}  
\bibliography{arxiv}  

\appendix
\newpage

\section{Data Pre-processing Details}
In the raw dataset we obtained, historical stock price time series and financial statement indicators can be utilized directly, while the primary construction lies in the pre-processing of news data. We provide more details on news preprocessing in this section, including text cleaning and news clustering.

\paragraph{Text Cleaning} Since the news dataset is scraped from web sources, they often contain a significant amount of information unrelated to the core content of the reports, such as irrelevant names, advertisements, and structural website text. We define these elements as “bias information.” Such information might mislead the analysis of LLM and consume computational resources inefficiently when processed by the models.

\begin{figure}[h]
\centering
\includegraphics[width=0.8\textwidth]{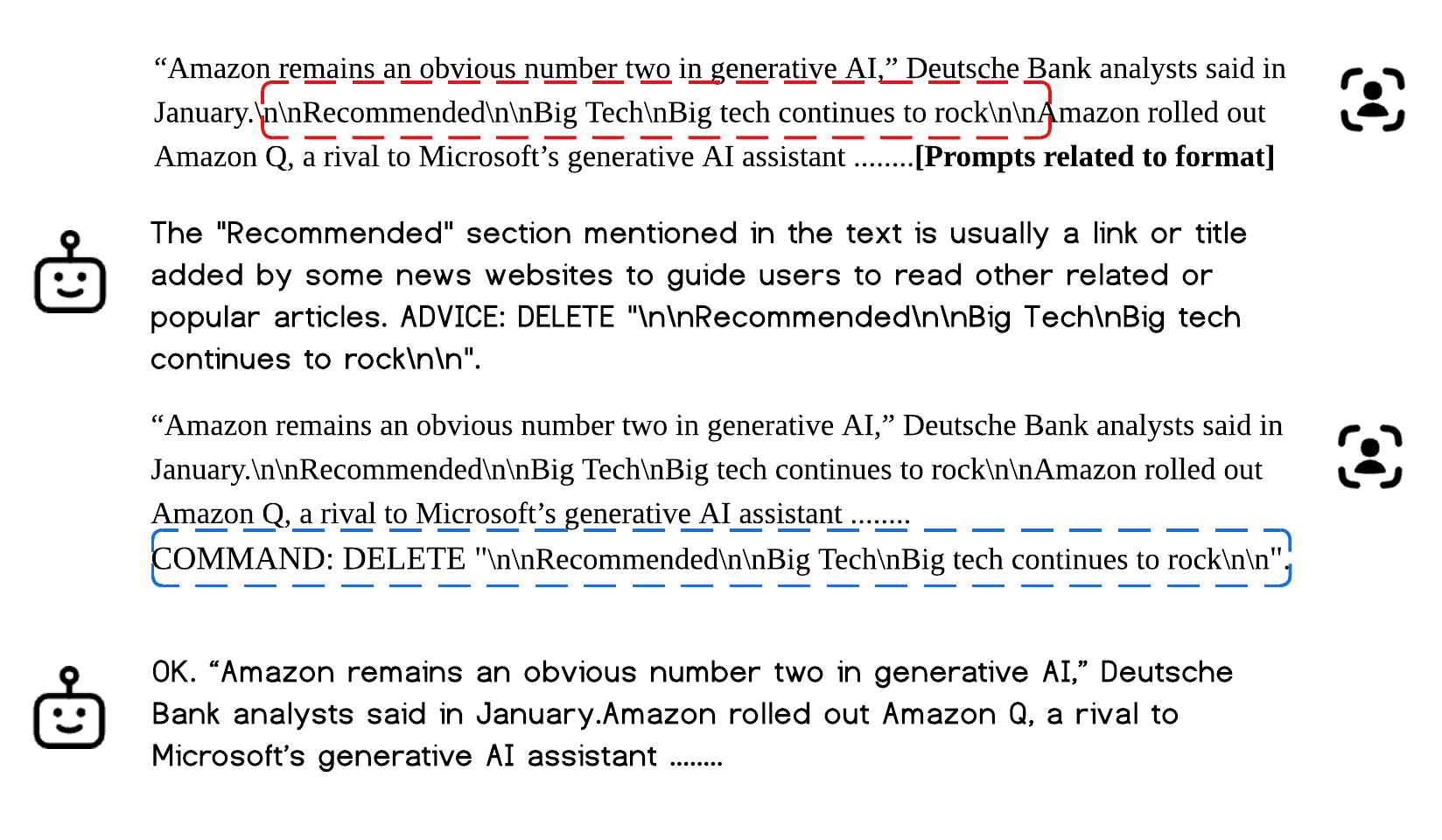}
\caption{\textbf{An example} of the process of pre-processing news data using LLM, with the \textcolor{red}{\textbf{red box}} displaying the ``bias information'' present in the news data and the \textcolor{blue}{\textbf{blue box}} indicating the modification commands given by LLM.}
\label{fig3}
\end{figure}

To address this issue, we employed regular expression matching to remove the known and fixed-pattern “bias information” initially. Subsequently, we conducted a self-reflection process equipped with LLM. Specifically, we feed each news article into the LLM, commanding it to identify any remaining “bias information” which haven't been removed and to suggest modifications, showing in \autoref{fig3}. The news articles, along with the suggested modifications, are then re-input into the LLM for correction. Through these steps, most of the “bias information” will be effectively eliminated from the dataset.

\paragraph{News Clustering} Another challenge in raw news data is the presence of multiple news articles about the target company on the same day or the inclusion of news compilations that cover more than just the target company. To handle this, we applied \textit{k-means} clustering to select the most representative news article from multiple same-day entries, thereby ensuring that the dataset contained only the most relevant and comprehensive information.

Even after pre-processing, the news dataset still contained proper nouns related to the target company, such as the names of individuals mentioned in news about executive appointments. These words may not be recognizable to LLM, potentially leading to hallucinations. To further address this issue, FinArena is equipped with RAG technology in the News Agent, which is specifically designed to improve the understanding and precision of the model by integrating relevant contextual information.

\end{document}